\documentclass[default]{sn-jnl}

\usepackage{graphicx}%
\usepackage{multirow}%
\usepackage{amsmath,amssymb,amsfonts}%
\usepackage{amsthm}%
\usepackage{wrapfig}
\usepackage{mathrsfs}%
\usepackage[title]{appendix}%
\usepackage{xcolor}%
\usepackage{textcomp}%
\usepackage{manyfoot}%
\usepackage{booktabs}%
\usepackage{algorithm}%
\usepackage{algorithmicx}%
\usepackage{algpseudocode}%
\usepackage{listings}%
\usepackage{subcaption}

\begin{document}
\title[]{Deformation Due to Non-planar Fault Movement in Fractional Maxwell Medium}

\author*[1]{\fnm{Pabita} \sur{Mahato}}\email{pabitamahato012@gmail.com,\\ https://orcid.org/0000-0002-6859-2140}

\author[2]{\fnm{Seema} \sur{Sarkar (Mondal)}}\email{ssarkarmondal.maths@nitdgp.ac.in}
\equalcont{These authors contributed equally to this work.}
\author[3]{\fnm{Subhash Chandra} \sur{Mondal}}\email{subhashch.mondal@gmail.com}
\equalcont{These authors contributed equally to this work.}
\affil[1,2]{\orgdiv{Department of Mathematics}, \orgname{National Institute of Technology Durgapur}, \orgaddress{\city{Durgapur}, \postcode{713209}, \state{West Bengal}, \country{India}}}
\affil[3]{\orgdiv{Department of Mathematics}, \orgname{P. K. H. N. Mahavidyalaya, Kanpur}, \orgaddress{\city{Howrah}, \postcode{711410}, \state{West Bengal}, \country{India}}}
\abstract{In earthquake-prone regions, the accumulation of geophysical stress during the aseismic period plays a critical role in determining which faults are more likely to be reactivated in future seismic events. In this model, we consider an infinite non-planar fault located in a viscoelastic half-space of a fractional Maxwell medium representing the lithosphere-asthenosphere system comprising three interconnected planar sections. The problem is formulated as a two-dimensional boundary value problem with discontinuities along the fault surface. A numerical solution is obtained using a Laplace transformation, fractional derivative, correspondence principle and Green's function technique. The outcomes are demonstrated graphically using appropriate model parameters. The computational findings highlight the significant influence of fault motion and geometry in shaping the displacement, stress and strain fields in the vicinity of the fault zone. A study has been carried out to investigate how non-planar faults influence displacement and the accumulation of stress and strain. Analysis of these results can provide insights into subsurface deformation and its impact on fault movement, which may contribute to the study of earthquake activity.}


\maketitle
\section{Introduction}\label{sec1}
A key question in studying the role of fault geometry in an earthquake cycle is how variations in fault geometry influence the stress distribution along the fault. Also, ground deformation that occurs in tectonically active areas between two significant seismic events is strongly influenced by fault activity and localized surface damage. Therefore, a detailed study of stress accumulation during this aseismic phase is essential for identifying which faults are more likely to experience future movement. Such analyses can be conducted by developing appropriate mathematical models that reflect the region's specific geological features and fault configurations. To develop our model, we have conducted literature surveys, some of which are listed below.

Major earthquakes typically occur on complex non-planar fault systems rather than simple planar faults. Various researchers have contributed to the modeling of spontaneous dynamic rupture along such irregular fault surfaces. For instance, early studies by Koller et al. (1992), Tada and Yamashita (1996), and Seelig and Gross (1997) utilized two-dimensional boundary integral techniques to simulate rupture propagation along non-planar faults. Expanding on this, Aochi et al. (2000) developed a numerical simulation approach to study spontaneous dynamic rupture propagation in a three-dimensional elastic medium. Cruz-Atienza et al. (2004) employed a staggered-grid finite-difference scheme to conduct two-dimensional modeling of seismic rupture processes. Later, Li et al. (2009) developed a 3-D viscoelastoplastic finite element model to simulate long-term, steady-state fault slip along conceptual restraining bends and crustal deformation around them. Oglesby and Mai (2012), Pelties et al. (2012), and Hisakawa et al. (2000) analysed three-dimensional non-planar faults by different numerical methods like the finite element method,  the discontinuous Galerkin method, the spectral element method and the boundary integral equation method, respectively. Mondal and Debsarma (2023) studied a non-planar strike-slip fault taking integer order derivatives. Folesky (2024) found that the earthquakes start in different ways depending on the location, smoother, weaker zones at the plate boundary, and deeper zones where the cause of quakes changes with depth. 

In theoretical models, the lithosphere-asthenosphere system is frequently idealized as a viscoelastic half-space to effectively represent the combined elastic and viscous responses that influence crustal deformation. This approach is particularly useful for understanding geophysical phenomena such as tectonic plate motion and post-seismic relaxation. The Maxwell model, which comprises a spring (elastic component) and a dashpot (viscous component) in series, allows for instantaneous elastic deformation followed by long-term viscous flow. Therefore, the Maxwell viscoelastic model may be a fundamental tool in understanding how the Earth’s crust and upper mantle respond to tectonic and surface loading. Furthermore, for a more realistic, non-linear representation that incorporates long-term memory effects, we employed the fractional Maxwell model.

Moreover, in viscoelastic theory, both stress and strain within a material are affected by immediate forces as well as by the history of deformation of the material. This time-dependent behaviour can be mathematically expressed through time integrals and therefore represented by fractional derivatives as they inherently capture the memory effects of viscoelastic materials. Using fractional derivatives in the constitutive equation enhances the ability to capture the intricate, time-dependent relationship between stress and strain more accurately (Samko et al. 1993, Deng and Morozov 2018). Moreover, the non-local nature of fractional derivatives makes them especially suitable for earthquake modeling, as they consider the distributed effects of stress and strain throughout the fault network (Wu et al. 2023). Studies have shown that fractional derivatives play a crucial role in altering the frictional resistance and stability characteristics of tectonic faults (El-Misiery and Ahmed, 2006). Pelap et al. (2018) and Tanekou et al. (2020) reported that the inclusion of a fractional-order derivative in the system dynamics can result in a shift from periodic behavior to a stable equilibrium condition. Mondal and Debnath (2021), Mahato et al. (2022), Mahato and Sarkar Mondal (2025) analyzed displacement stress and strain of different types of planar faults in a viscoelastic half-space using fractional derivatives. Romnet et al. (2024) isolate the influence of fault non-planarity by expanding the boundary integral equation, effectively separating it from the response of a planar fault. Zielke and Mai (2025) explore the influence of fault geometry and variations in strength on long-term fault behaviour, specifically examining how these factors impact the magnitude–frequency relationship, the time intervals between earthquakes, and the size of the largest possible events. As far as we are aware, limited research has been conducted on non-planar faults embedded in a viscoelastic medium using fractional derivative approaches.

Considering these factors, this study examines a non-planar strike-slip fault represented by three planar segments situated in a viscoelastic half-space of a fractional Maxwell medium. Analytical solutions for displacement, stress, and strain have been evaluated both prior to and after the fault movement by employing the Laplace transform of the fractional derivative, the Green's function technique, and the correspondence principle. The organization of the paper is as follows: Section \ref{sec2} outlines the development of the model. Section \ref{sec3} presents the calculations to derive displacement, stress, and strain in both pre- and post-fault movement scenarios. Section \ref{sec4} provides graphical interpretations of the results and includes a comparative study with an integer-order model. Finally, Section \ref{sec5} includes concluding remarks and future scope, followed by declarations and references. 
         
\section{Mathematical model}\label{sec2}
This study examines a theoretical framework within the lithosphere-asthenosphere system, assuming an infinite, non-planar fault $F$ situated in a fractional Maxwell viscoelastic medium. The fault is considered surface-breaking with strike-slip motion. The surface of this non-planar fault deviates from a simple, flat plane, and it involves various bends. The non-planar fault is embedded with three planar parts of width $l_1$, $l_2$ and $l_3$. Each segment is tilted at a specific angle relative to the horizontal plane.
\subsection{Formulation}
To define the fault geometry, a Cartesian coordinate system $(y_1, y_2, y_3)$ is used, where the $y_1$-axis runs along the fault's strike direction, the $y_3$-axis extends vertically downward, representing the depth, and the $y_2$-axis is perpendicular to the $y_1y_3$ plane. In this setup, the free surface corresponds to $y_3 = 0$, while the region $y_3 \geq 0$ represents the viscoelastic half-space. Since the fault is an infinite fault, we have considered the cross-section of the model by the plane $y_1=0$. The segments $AB$, $BC$, $CD$ of widths $l_1$, $l_2$, $l_3$ inclined at angles $\theta_1$, $\theta_2$ and $\theta_3$ respectively with the Earth's surface. The figure \ref{figure1} represents a cross-sectional view of the model along the plane $y_1 = 0$. For each planar part of the fault we have considered three more coordinate systems $(y_1', y_2', y_3')$ for $AB$, $(y_1'', y_2'', y_3'')$ for $BC$ and $(y_1''', y_2''', y_3''')$ for $CD$ where $y_1'$, $y_1''$ and $y_1'''$ are parallel to $y_1$ and are related as follows:  
\begin{figure}
      \centering     \includegraphics[width=0.6\linewidth]{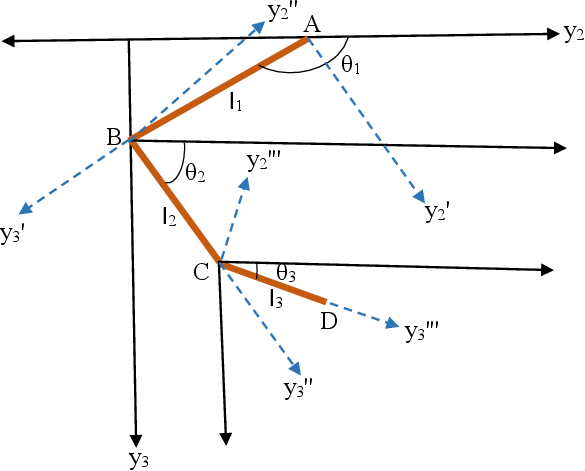}
      \caption{Schematic diagram of the model on the plane $y_1=0$.}
      \label{figure1}
  \end{figure}
\begin{equation}
    \begin{cases}   y_2=y_2'\sin\theta_1+y_3'\cos\theta_1-l_1\cos\theta_1\\y_3=-y_2'\cos\theta_1+y_3'\sin\theta_1
    \end{cases}
    \label{equation1}
\end{equation}
\begin{equation}
    \begin{cases}
y_2=y_2''\sin\theta_2+y_3''\cos\theta_2\\
y_3=-y_2''\cos\theta_2+y_3''\sin\theta_2+l_1\sin\theta_1
    \end{cases}
    \label{equation2}
\end{equation}
and
\begin{equation}
    \begin{cases}
y_2=y_2'''\sin\theta_3+y_3'''\cos\theta_3+l_2\cos\theta_2\\
y_3=-y_2'''\cos\theta_3+y_3'''\sin\theta_3+l_1\sin\theta_1+l_2\sin\theta_2
    \end{cases}
    \label{equation3}
\end{equation}\\
Let us assume that $u_1$, $u_2$ and $u_3$ be the displacement components along $y_1$, $y_2$, $y_3$ respectively; $\tau_{11}$, $\tau_{12}$, $\tau_{13}$, $\tau_{22}$, $\tau_{23}$, $\tau_{33}$ be the stress components and $e_{11}$, $e_{12}$, $e_{13}$, $e_{22}$, $e_{23}$, $e_{33}$ be strain components. Since this fault is an infinite strike-slip fault, the associated displacement, stress and strain components are $u_1, \tau_{12}, \tau_{13}, e_{12}, e_{13}$ (Mukhopadhyay et al. 1980) and they are all independent of $y_1$.\\\\
Our analysis begins at $t=0$, assuming no initial fault slip within the medium.  However, slow aseismic deformation continues to take place within the medium. Shear stresses in the medium arise due to various tectonic forces, including mantle convection, weight of overlying material, and other geological processes. The resulting displacement and stress must then adhere to specific constitutive laws, boundary conditions for $t\geq0$, and initial conditions defined at $t=0$, as outlined below.
\subsection{Constitutive equation}\label{sec2.1}
The stress-strain relations of the model are as given below (Debnath 2003):
\begin{eqnarray}
  \frac1{\eta}(\tau_{12})+\frac{1}{\mu}\,{_0D^{\alpha}_t} (\tau_{12})= {_0D^{\alpha}_t} \left(\frac{\partial u_1}{\partial y_2}\right)
    \label{equation4}
\end{eqnarray}
\begin{eqnarray}
  \frac1{\eta}(\tau_{13})+\frac{1}{\mu}\,{_0D^{\alpha}_t} (\tau_{13})= {_0D^{\alpha}_t} \left(\frac{\partial u_1}{\partial y_3}\right)
    \label{equation5}
\end{eqnarray}
for $-\infty<y_2<\infty$, $y_3\geq0$, $t\geq 0$. Here, $\eta$ and $\mu$ represent the effective viscosity and effective rigidity of the viscoelastic medium, respectively. The operator $_0D_t^{\alpha}$ denotes the fractional derivative of order $\alpha$ ($\alpha\in (0, 1]$), defined over the interval from 0 to $t$. In this context, the Caputo definition of the fractional derivative (Caputo 1969) is considered, which is defined as
\begin{equation*}
   _0 D_t^{\alpha}(f(t))=\frac{1}{\Gamma(1-\alpha)}\int_0^t \frac{f'(\xi)}{(t-\xi)^{\alpha}}d\xi
\end{equation*}
\subsection{Stress equation of motion}\label{sec2.2}
The inertial forces are expected to be negligible because of the slow and quasi-static deformation in the aseismic period, as discussed by Mukhopadhyay et al. (1980). Additionally, as the body forces remain uniform throughout the medium relative to the initial stress state, their variation can be regarded as negligible. As a result, the stress equation governing strike-slip motion can be simplified accordingly:
\begin{eqnarray}
   \frac{\partial \tau_{12}}{\partial y_2}+\frac{\partial \tau_{13}}{\partial y_3}=0
    \label{equation6}
\end{eqnarray}
for $t\geq 0$, $-\infty<y_2<\infty$, $y_3\geq0$.
\subsection{Boundary conditions}\label{sec2.3}
As $|y_2|\to\infty$, the stress component $\tau_{12}$ approaches the stress $\tau_{\infty}(t)$, which is caused by tectonic forces within the lithosphere-asthenosphere system. This stress evolves over time and is given by $\tau_{\infty}(t)=\tau_{\infty}(0)(1+kt)$ where $\tau_{\infty}(0)$ is the initial tectonic stress at $t=0$ and $k$ is very small positive constant, i.e.,
\begin{eqnarray}
    \tau_{12}\to\tau_{\infty}(t)=\tau_{\infty}(0)(1+kt),\, k>0 \mbox{ for } |y_2|\to\infty, y_3\geq0, t\geq0
    \label{equation7}
\end{eqnarray}
To a practical approximation, there is no stress transferred between the Earth's surface and the atmosphere. Therefore, on the Earth's surface, we can take the boundary condition on $\tau_{13}$ as below:
\begin{eqnarray}
    \tau_{13}=0\mbox{ on } y_3=0, \mbox{ for } -\infty\leq y_2\leq\infty, t\geq0
    \label{equation8}
\end{eqnarray}
Also, as $y_3\to \infty$,
\begin{eqnarray}
    \tau_{13}\to0, \mbox{ for } -\infty\leq y_2\leq\infty, t\geq0
    \label{equation9}
\end{eqnarray}
\subsection{Initial conditions}\label{sec2.4}
As there will be impact of the initial field, the initial values of displacement, stress and strain components are assumed as ${(u_1)}_0$, ${(\tau_{12})}_0, {(\tau_{13})}_0$,  ${(e_{12})}_0, {(e_{13})}_0$ respectively. All of them are constants or functions of $y_2$, $y_3$, independent of $t$ and satisfies all the relations (\ref{equation4}) - (\ref{equation9}).
\section{Solutions}\label{sec3} 
 Now, we will solve the above time-dependent boundary value problem with the help of the Laplace transformation to reduce viscoelastic problems to elastic ones. Differentiating (\ref{equation4}) and (\ref{equation5}) partially with respect to $y_2$ and $y_3$, respectively and adding them (Omitting the first and higher-order derivatives of $u_1$, as their magnitudes are negligible for $0 < \alpha \leq 1$), we get the governing equation as
\begin{equation}
    \nabla^2 U=0 \mbox{ where } U=u_1-{(u_1)}_0
    \label{equation10}
\end{equation}
Taking Laplace transformation of (\ref{equation10}) we get
\begin{equation}
    \nabla^2 \overline{U}=0 \mbox{ where } \overline U=\overline{u_1}-\frac{{(u_1)}_0}s
    \label{equation11}
\end{equation}
where $\overline{U}$, $\overline{u_1}$ are Laplace transforms of $U$ and $u_1$ respectively with respect to time $t$ and $s$ is the Laplace transform variable. 
\subsection{Displacement, stress \& strain before fault movement}
Let us assume a trial solution of equation (\ref{equation11}) as
\begin{equation}
    \nabla^2 \overline{U}=Ay_2+By_3 \mbox{ where } \overline U=\overline{u_1}-\frac{{(u_1)}_0}s
    \label{equation12}
\end{equation}
where $A, B$ are constants or functions of $s$ in transformed domain but independent of $y_2$ and $y_3$.

Let $\overline{\tau_{12}}$, $\overline{\tau_{13}}$ be the Laplace transforms of $\tau_{12}, \tau_{13}$ respectively. Taking Laplace transformation of the relations (\ref{equation4}) to (\ref{equation9}) (omitting both the first and all higher-order derivatives of ${u_1}, {\tau_{12}}, {\tau_{13}}$ as they are negligibly small for $0<\alpha\leq 1$), we get
\begin{equation}
     \left(1+\frac{\eta}{\mu}s^{\alpha}\right)\overline{\tau_{12}}-\frac{\eta}{\mu}s^{\alpha-1}{(\tau_{12})}_0=\eta s^{\alpha-1}\frac{\partial }{\partial y_2}\left(s\overline {u_1}-{(u_1)}_0\right)
    \label{equation13}
\end{equation}
\begin{equation}
     \left(1+\frac{\eta}{\mu}s^{\alpha}\right)\overline{\tau_{13}}-\frac{\eta}{\mu}s^{\alpha-1}{(\tau_{13})}_0=\eta s^{\alpha-1}\frac{\partial }{\partial y_2}\left(s\overline {u_1}-{(u_1)}_0\right)
    \label{equation14}
\end{equation}
\begin{eqnarray}
    \frac{\partial \overline{\tau_{12}}}{\partial y_2}+\frac{\partial \overline{\tau_{13}}}{\partial y_3}=0
    \label{equation15}
\end{eqnarray}
\begin{eqnarray}
        \overline{\tau_{12}}\to\tau_{\infty}(0)\left(\frac{1}{s}+\frac{k}{s^2}\right)
    \label{equation16}
\end{eqnarray}
\begin{equation}
\begin{cases}
    \overline{\tau_{13}}= 0 \mbox{ on } y_3=0\\
        \overline{\tau_{13}}\to 0 \mbox{ as } y_3\to\infty
\end{cases}
\label{equation17}
\end{equation}
Using the trial solution (\ref{equation12}) and boundary conditions of transformed domain (\ref{equation16}), (\ref{equation17}) in (\ref{equation13}) and (\ref{equation14}) we get the unknowns $A$ and $B$ as
\begin{eqnarray}
    \begin{cases}
        A=\frac{{\tau}_{\infty}(0)}{\eta}\left(\frac1{s^{\alpha+1}}+\frac k{s^{\alpha+2}}+\frac{\eta k}{\mu s^2}\right)\\
        B=0   
    \end{cases}
    \label{equation18}
\end{eqnarray}
By substituting (\ref{equation18}) into (\ref{equation12}), and then incorporating the resulting expression into (\ref{equation13}) and (\ref{equation14}) and finally there inverse Laplace transformation gives
\begin{eqnarray}
    \begin{cases}
        u_1={(u_1)}_0+\frac{\tau_{\infty}(0)y_2}{\eta}\left\{\frac{t^{\alpha}}{\Gamma(\alpha+1)}+\frac{kt^{\alpha+1}}{\Gamma(\alpha+2)}+\frac{\eta kt}{\mu}\right\}\\
        \tau_{12}=\left({(\tau_{12})}_0-\tau_{\infty}(0)\right)E_{\alpha}\left(-\frac{\mu}{\eta}t^{\alpha}\right)+\tau_{\infty}(0)(1+kt)\\
        \tau_{13}={(\tau_{13})}_0E_{\alpha}\left(-\frac{\mu}{\eta}t^{\alpha}\right)\\
        e_{12}=\frac{\partial u_1}{\partial y_2}={(e_{12})}_0+\frac{\tau_{\infty}(0)}{\eta}\left\{\frac{t^{\alpha}}{\Gamma(\alpha+1)}+\frac{kt^{\alpha+1}}{\Gamma(\alpha+2)}+\frac{\eta kt}{\mu}\right\}\\ 
        e_{13}=\frac{\partial u_1}{\partial y_3}={(e_{13})}_0
    \end{cases}
    \label{equation19}
\end{eqnarray}
where $\Gamma$ is the Gamma function.\\
The expression for $\tau_{12}$ in equation (\ref{equation19}) indicates that the resulting stress in the medium depends on $y_2, y_3$ and increases with time $t$. Considering the rheological behaviour of viscoelastic materials near the fault, we can assume that the medium can sustain only a limited amount of stress, denoted by a critical threshold $\tau_c$. It is further assumed that after a specific time $T$ (say), referred to as the critical time, the accumulated stress in the fault's vicinity surpasses this threshold, leading to slip or movement along the fault.
\subsection{Solution after the commencement of fault movement}
After the critical time $T$, there is a transition from aseismic to seismic state which remains for a small duration of time ($t_{\epsilon}$ say), which usually lasts for a few seconds. During this time, seismic disturbances occur and then gradually diminish, allowing the medium to transit back to a quasi-static and aseismic state. We now examined this model at the time, immediately after the re-establishment of the new aseismic state in the viscoelastic medium.\\
In this case, the displacement, stress and strain components satisfy all the relations (\ref{equation4}) to (\ref{equation9}) for a new time variable $t_1\geq0$ where $t_1=t-T$, excluding the boundary condition (\ref{equation7}). Additionally, the displacement components $u_1$ exhibit a discontinuity due to the dislocation across the fault $F$, which is represented as follows:
\begin{eqnarray}
    {[u_1]}_{F}=U(t_1)f(\xi)H(t_1),\, t_1\geq0
    \label{equation20}
\end{eqnarray}
Here, the term $U(t_1)$ describes the dislocation occurring at the top end of the fault located on the free surface. $H(t_1)$ stands for the Heaviside step function, which captures the onset of this dislocation over time. The function $f(\xi)$ defines how dislocation varies with depth, where $\xi$ is measured downward along the fault line at $y_1 = 0$. The notation ${[u_1]}_{F}$ represents the discontinuity or abrupt change in the displacement component $u_1$ across the fault $F$ as a result of fault slip and is defined by
\begin{eqnarray}
    {[u_1]}_{F}=\lim_{y_2'\to0^+}(u_1)-\lim_{y_2'\to0^-}(u_1)
    \label{equation21}
\end{eqnarray}
where the relations between $(y_2, y_3)$ and $(y_2', y_3')$ are given in (\ref{equation1}).\\
Also, stress $\tau_{12}$ satisfy new boundary condition
\begin{eqnarray}
     {\tau_{12}}\to0 \mbox{ as } |y_2|\to \infty,\, y_3\geq 0,\, t_1> 0
     \label{equation22}
\end{eqnarray}
Now, the final solution can be taken in the form
\begin{equation}
    \begin{cases}
    u_1={(u_1)}_1+{(u_1)}_2\\
    \tau_{12}={(\tau_{12})}_1+{(\tau_{12})}_2\\
    \tau_{13}={(\tau_{13})}_1+{(\tau_{13})}_2\\
    e_{12}={(e_{12})}_1+{(e_{12})}_2\\
    e_{13}={(e_{13})}_1+{(e_{13})}_2
    \end{cases}
    \label{equation23}
\end{equation}
where ${(u_1)}_1$, ${(\tau_{12})}_1, {(\tau_{13})}_1$, ${(e_{12})}_1, {(e_{13})}_1$ are displacement, stress and strain components before the movement of the fault and are same as given in (\ref{equation19}) while ${(u_1)}_2$, ${(\tau_{12})}_2, {(\tau_{13})}_2$, ${(e_{12})}_2, {(e_{13})}_2$ are respectively those after the movement of the fault which satisfies the additional conditions (\ref{equation20}) and (\ref{equation22}). To find these components, we have to take the Laplace transformation of the constitutive equations (\ref{equation4}) and (\ref{equation5}) for $t_1\geq 0$, and we get
\begin{eqnarray}
     {(\overline{\tau_{12}})}_2=\frac{\mu s^{\alpha}}{s^{\alpha}+\frac{\mu}{\eta}}\frac{\partial {(\overline {u_1})}_2}{\partial y_2}
     \label{equation24}
     \end{eqnarray}
     \begin{eqnarray}
     {(\overline{\tau_{13}})}_2=\frac{\mu s^{\alpha}}{s^{\alpha}+\frac{\mu}{\eta}}\frac{\partial {(\overline {u_1})}_2}{\partial y_3}
     \label{equation25}
     \end{eqnarray}
where $s$ is the Laplace transform variable for the new time coordinate $t_1$.\\

We take the dislocation of the fault $F$ as $U(t_1)=Vt_1$ where $V$ is the constant creep velocity of the fault $F$. Substituting this in (\ref{equation20}) and taking Laplace transformation of (\ref{equation20}) and (\ref{equation22}) we get
\begin{eqnarray}
{[\overline{u_1}]}_{F_1}=\frac{V}{s^2}f(\xi)
    \label{equation26}
    \end{eqnarray}
    \begin{eqnarray}  
    {(\overline{\tau_{12}})}_2\to 0 \mbox{ as } |y_2|\to \infty, y_3\geq 0
    \label{equation27}
\end{eqnarray}
Now, this boundary value problem is solved with the help of Green's function technique (Maruyama, 1966) and the correspondence principle (Rybiki, 1968). Let $P(\xi_1, \xi_2, \xi_3)$ be any point on the fault and $Q(y_1, y_2, y_3)$ be any field point on the half-space, then the displacement component is given by
\begin{equation}
    \overline{(u_1)}_2=\int_F\overline{[u_1]}_2(s)\{G_{12}(P, Q)d\xi_3-G_{13}(P, Q)d\xi_2\}
    \label{equation28}
\end{equation}
where 
\begin{equation}
    G_{12}=\frac{1}{2\pi}\left[\frac{y_2-\xi_2}{(y_2-\xi_2)^2+(y_3-\xi_3)^2}+\frac{y_2-\xi_2}{(y_2-\xi_2)^2+(y_3+\xi_3)^2}\right]
    \label{equation29}
\end{equation}
and
\begin{equation}
    G_{13}=\frac{1}{2\pi}\left[\frac{y_3-\xi_3}{(y_2-\xi_2)^2+(y_3-\xi_3)^2}-\frac{y_3+\xi_3}{(y_2-\xi_2)^2+(y_3+\xi_3)^2}\right]
    \label{equation30}
\end{equation}
To find the displacement for the planar part $AB$, we will transform the fault point $P(\xi_1, \xi_2, \xi_3)$ to $(\xi_1', \xi_2', \xi_3')$ as follows
\begin{equation}
\begin{cases}
    \xi_2=\xi_2'\sin\theta_1+\xi_3'\cos\theta_1-l_1\cos\theta_1\\
    \xi_2=-\xi_2'\cos\theta_1+\xi_3'\sin\theta_1
    \end{cases}
    \label{equation31}
\end{equation}
such that on the fault, $\xi_2'=0$.\\
Use of (\ref{equation29}), (\ref{equation30}) and (\ref{equation31}) in (\ref{equation28}) and the its inverse Laplace transformation gives dislocation as
\begin{equation}
    {\{{(u_1)}_2\}}_{AB}=\frac{V}{2\pi s^2}\Phi_1(y_2, y_3)
     \label{equation32}
\end{equation}
where
\begin{multline}
   \Phi_1(y_2, y_3)=\int_0^{l_1}f(\xi_3')\Bigg[\frac{y_2\sin\theta_1-y_3\cos\theta_1+l_1\sin\theta_1\cos\theta_1}{(y_2-(\xi_3'-l_1)\cos\theta_1)^2+(y_3-\xi_3'\sin\theta_1)^2}\cr+\frac{y_2\sin\theta_1+y_3\cos\theta_1+l_1\sin\theta_1\cos\theta_1}{(y_2-(\xi_3'-l_1)\cos\theta_1)^2+(y_3+\xi_3'\sin\theta_1)^2}\Bigg]d\xi_3'
    \label{equation33}
\end{multline}
Similarly, we can find the dislocation components for the parts $BC$ and $CD$ as follows:
\begin{equation}
    {\{{(u_1)}_2\}}_{BC}=\frac{V}{2\pi s^2}\Psi_1(y_2, y_3)
     \label{equation34}
\end{equation}
and
\begin{equation}
    {\{{(u_1)}_2\}}_{CD}=\frac{V}{2\pi s^2}\chi_1(y_2, y_3)
     \label{equation35}
\end{equation}
where
\begin{multline}
   \Psi_1(y_2, y_3)=\int_{l_1\frac{\sin\theta_1}{\sin\theta_2}}^{l_1\frac{\sin\theta_1}{\sin\theta_2}+l_2}f(\xi_3'')\Bigg[\frac{y_2\sin\theta_2-y_3\cos\theta_2+l_1\sin\theta_1\cos\theta_2}{(y_2-\xi_3''\cos\theta_2)^2+(y_3-\xi_3''\sin\theta_2-l_1\sin\theta_1)^2}\cr+\frac{y_2\sin\theta_2+y_3\cos\theta_2+l_1\sin\theta_1\cos\theta_2}{(y_2-\xi_3''\cos\theta_2)^2+(y_3+\xi_3''\sin\theta_2+l_1\sin\theta_1)^2}\Bigg]d\xi_3''
    \label{equation36}
\end{multline}
and
\begin{multline}
   \chi_1(y_2, y_3)=\int_{l_2\frac{\sin\theta_2}{\sin\theta_3}}^{l_2\frac{\sin\theta_2}{\sin\theta_3}+l_3}f(\xi_3''')\cr\Bigg[\frac{y_2\sin\theta_3-y_3\cos\theta_3+l_1\sin\theta_1\cos\theta_3+l_2\sin(\theta_2-\theta_3)}{(y_2-\xi_3'''\cos\theta_3-l_2\cos\theta_2)^2+(y_3-\xi_3'''\sin\theta_3-l_1\sin\theta_1-l_2\sin\theta_2)^2}\cr+\frac{y_2\sin\theta_3+y_3\cos\theta_3+l_1\sin\theta_1\cos\theta_3+l_2\sin(\theta_2-\theta_3)}{(y_2-\xi_3'''\cos\theta_3-l_2\cos\theta_2)^2+(y_3+\xi_3'''\sin\theta_3+l_1\sin\theta_1+l_2\sin\theta_2)^2}\Bigg]d\xi_3'''
    \label{equation37}
\end{multline}
Hence, the dislocation component after the movement of the fault in the transformed domain is given by
\begin{equation}
    {(\overline{u_1})}_2=\frac{V}{2\pi s^2}[\Phi_1(y_2, y_3)+\Psi_1(y_2, y_3)+\chi_1(y_2, y_3)]
    \label{equation38}
\end{equation}
and its inverse Laplace transformation gives
\begin{equation}
    {(u_1)}_2=\frac1{2\pi}Vt_1H(t_1)[\Phi_1(y_2, y_3)+\Psi_1(y_2, y_3)+\chi_1(y_2, y_3)]
    \label{equation39}
\end{equation}
Substituting (\ref{equation38}) in (\ref{equation24}) and (\ref{equation25}), and taking the inverse Laplace transformation, we get the stress components as
\begin{equation}
    \begin{cases}
        {(\tau_{12})}_2=\frac1{2\pi}\mu VH(t_1)\left[t_1+\frac{\eta}{\mu}\left(1-E_{\alpha}\left(-\frac{\mu}{\eta}t_1^{\alpha}\right)\right)\right]\left[\Phi_2(y_2, y_3)+\Psi_2(y_2, y_3)+\chi_2(y_2, y_3)\right]\\
        {(\tau_{13})}_2=\frac1{2\pi}\mu VH(t_1)\left[t_1+\frac{\eta}{\mu}\left(1-E_{\alpha}\left(-\frac{\mu}{\eta}t_1^{\alpha}\right)\right)\right]\left[\Phi_3(y_2, y_3)+\Psi_3(y_2, y_3)+\chi_3(y_2, y_3)\right]
        \label{equation40}
    \end{cases}
\end{equation}
where $\Phi_2=\frac{\partial \Phi_1}{\partial y_2}$, $\Phi_3=\frac{\partial \Phi_1}{\partial y_3}$; $\Psi_2=\frac{\partial \Psi_1}{\partial y_2}$, $\Psi_3=\frac{\partial \Psi_1}{\partial y_3}$, $\chi_2=\frac{\partial \chi_1}{\partial y_2}$, $\chi_3=\frac{\partial \chi_1}{\partial y_3}$.\\
Using the formula $e_{12}=\frac{\partial u_1}{\partial y_2}$ and $e_{13}=\frac{\partial u_1}{\partial y_3}$ we get the strain components as
\begin{equation}
    \begin{cases}
        {(e_{12})}_2=\frac1{2\pi}Vt_1H(t_1)[\Phi_2(y_2, y_3)+\Psi_2(y_2, y_3)+\chi_2(y_2, y_3)]\\
        {(e_{13})}_2=\frac1{2\pi}Vt_1H(t_1)[\Phi_3(y_2, y_3)+\Psi_3(y_2, y_3)+\chi_3(y_2, y_3)]
    \end{cases}
    \label{equation41}
\end{equation}
Finally, the complete solution for displacement, stress and strain is obtained by substitution of (\ref{equation39}), (\ref{equation40}), (\ref{equation41}) and (\ref{equation19}) in (\ref{equation23}) as given below.
\begin{multline}
    \begin{cases}
        u_1={(u_1)}_0+\frac{\tau_{\infty}(0)y_2}{\eta}\left\{\frac{t^{\alpha}}{\Gamma(\alpha+1)}+\frac{kt^{\alpha+1}}{\Gamma(\alpha+2)}+\frac{\eta kt}{\mu}\right\}\cr\hspace{0.4cm}+\frac1{2\pi}Vt_1H(t_1)[\Phi_1(y_2, y_3)+\Psi_1(y_2, y_3)+\chi_1(y_2, y_3)]\\
        \tau_{12}=\left({(\tau_{12})}_0-\tau_{\infty}(0)\right)E_{\alpha}\left(-\frac{\mu}{\eta}t^{\alpha}\right)+\tau_{\infty}(0)(1+kt)\cr\hspace{1cm}+\frac1{2\pi}\mu VH(t_1)\left[t_1+\frac{\eta}{\mu}\left(1-E_{\alpha}\left(-\frac{\mu}{\eta}t_1^{\alpha}\right)\right)\right]\cr\hspace{1cm}\left[\Phi_2(y_2, y_3)+\Psi_2(y_2, y_3)+\chi_2(y_2, y_3)\right]\\
        \tau_{13}={(\tau_{13})}_0E_{\alpha}\left(-\frac{\mu}{\eta}t^{\alpha}\right)+\frac1{2\pi}\mu VH(t_1)\left[t_1+\frac{\eta}{\mu}\left(1-E_{\alpha}\left(-\frac{\mu}{\eta}t_1^{\alpha}\right)\right)\right]\cr\hspace{1cm}\left[\Phi_3(y_2, y_3)+\Psi_3(y_2, y_3)+\chi_3(y_2, y_3)\right]\\
        e_{12}={(e_{12})}_0+\frac{\tau_{\infty}(0)}{\eta}\left\{\frac{t^{\alpha}}{\Gamma(\alpha+1)}+\frac{kt^{\alpha+1}}{\Gamma(\alpha+2)}+\frac{\eta kt}{\mu}\right\}\cr\hspace{1cm}+\frac1{2\pi}Vt_1H(t_1)[\Phi_2(y_2, y_3)+\Psi_2(y_2, y_3)+\chi_2(y_2, y_3)]\\ 
        e_{13}={(e_{13})}_0+\frac1{2\pi}Vt_1H(t_1)[\Phi_3(y_2, y_3)+\Psi_3(y_2, y_3)+\chi_3(y_2, y_3)]
    \end{cases}
     \label{equation43}
\end{multline}
where $\Phi_i, \Psi_i, \chi_i$, $i=1, 2, 3$ are as mentioned above. 
\section{Numerical Computation}\label{sec4}
Now, we investigate how fault movement influences displacement, stress, and strain for a range of creep velocities and orders of fractional derivative. The parameters used in this analysis are so chosen that they lie within their feasible range, and they are guided by prior studies that utilized empirical data from natural geological settings, as referred to in our discussion.\\
The rigidity value $(\mu)$ of common crustal materials, like granite or basalt, typically fall within the range of $10^{10}-10^{11}$ $N/m^2$ (Turcotte and Gerald 2002) and by the observations of dense geodetic, viscosities ($\eta$) in lower crust ranges between $10^{16}-10^{21}$ Pa·s (Bischoff and Flesch 2018) and that of in upper mantle is $10^{20}-10^{21}$ Pa·s (Ge et al. 2022). Keeping this in view, the value of rigidity $(\mu)$ and viscosity $(\eta)$are taken as follows\\\\
$\mu=3.5\times 10^{10}$ $N/m^2$ and $\eta=5\times 10^{19}$ Pa·s. (Mondal and Debsarma 2023)\\\\
The width of the fault lies in the range of $5$ to $15$ km (Mahato et al. 2022). In our model, the fault consists of three interconnected segments of different widths. Following the article (Mondal and Debsarma 2023), we consider\\\\
$l_1=5$ km, $l_2=3$ km, $l_3=2.5$ km.\\\\
The inclination of each planar part is taken in the range of $0<\theta\leq90^{\circ}$, and for the angle greater than $90^{\circ}$, we consider $\theta_1=180^{\circ}-\theta'$ such that $0<\theta'\leq90^{\circ}$. In this case, we consider the angles $\theta_1=140^{\circ}, \theta_2=60^{\circ}, \theta_3=30^{\circ}$ for numerical computation.\\\\
The fractional derivative order $(\alpha)$ lies within the range 
$0<\alpha\leq 1$. For our observations, we consider the values of $\alpha$ as 0.5, 0.9, and 1.\\\\
Titus et al. (2006) observed that the fault creep velocity lies in a range of $21$ and $41$ mm/year. So, for numerical computation, we take the creep velocities as $V = 0.02, 0.03, 0.04$ m/year.\\\\
The boundary condition of $\tau_{12}$ (equation (\ref{equation7})) is chosen such that the tectonic force $\tau_{\infty}(t)$ gradually increases linearly. Then, following Mahato et al. (2022), we take $\tau_{12}\to\tau_{\infty}(0)(1+kt)$ where $k$ is very small, taken as $k=10^{-9}$.\\\\
Initial stress ${(\tau_{12})}_0=20$ bar and $\tau_{\infty}(0)=50$ bar.\\\\ 
Following Mondal and Debsharma (2023), we take the critical stress $(\tau_c=200)$ bar, and we obtained that, for $\alpha=0.5$, the stress component $\tau_{12}$ (from (\ref{equation19})) exceed this critical value of stress after approx. $114.01$ years. Therefore, the critical time at which the fault's movement takes place is $T=114.01$ years. This critical time can be different for different values of critical stress and for different values of $\alpha$.\\\\
The dislocation functions $f(\xi)$ for each parts of the fault $F$ are taken as below
\begin{eqnarray}
    \begin{cases}
        f(\xi)&=\frac12\left[2-2\left(\frac{\xi}{l_1}\right)^3+\left(\frac{\xi}{l_1}\right)^4\right], \quad \mbox{for } AB\\
        &=\frac16\left[\sqrt3-(\sqrt3-\sqrt2)\left(\frac{\xi-l_1}{l_2}\right)^2\right]^2, \quad \mbox{for } BC\\
        &=\frac16\left[1-\frac56\left(\frac{\xi-l_1-l_2}{l_3}\right)^2+\frac13\left(\frac{\xi-l_1-l_2}{l_3}\right)^3\right], \quad \mbox{for } CD\\
    \end{cases}
\end{eqnarray}
 These functions fulfil all the necessary conditions, as explained by Mondal and Debsharma (2023), ensuring bounded displacement, stress and strain throughout its domain.\\

\begin{figure}[ht]
\begin{subfigure}{.5\textwidth}
  \centering
  \includegraphics[width=1\linewidth]{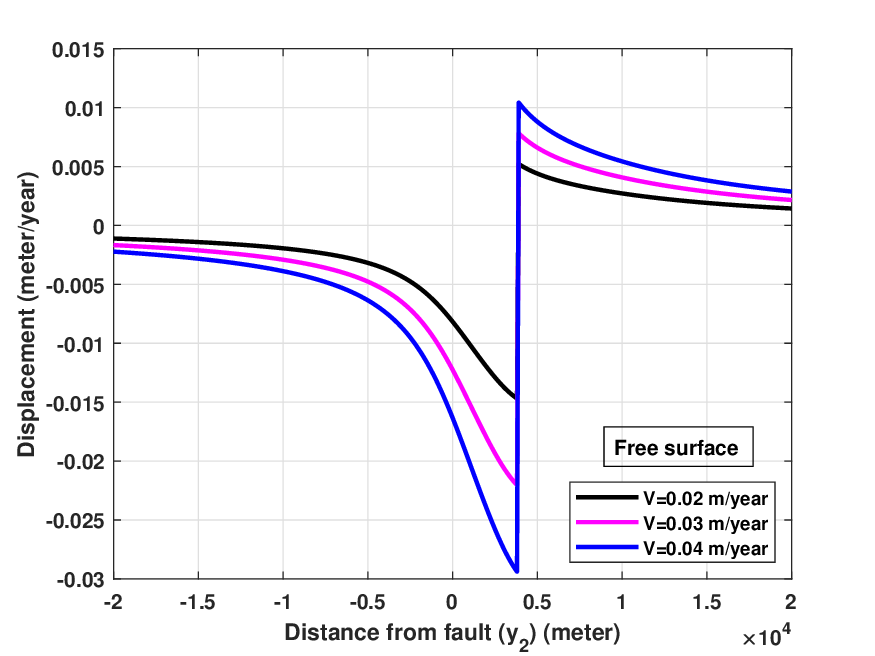}
  \caption{}
  \label{figure2a}
\end{subfigure}
\begin{subfigure}{.5\textwidth}
  \centering
  \includegraphics[width=1\linewidth]{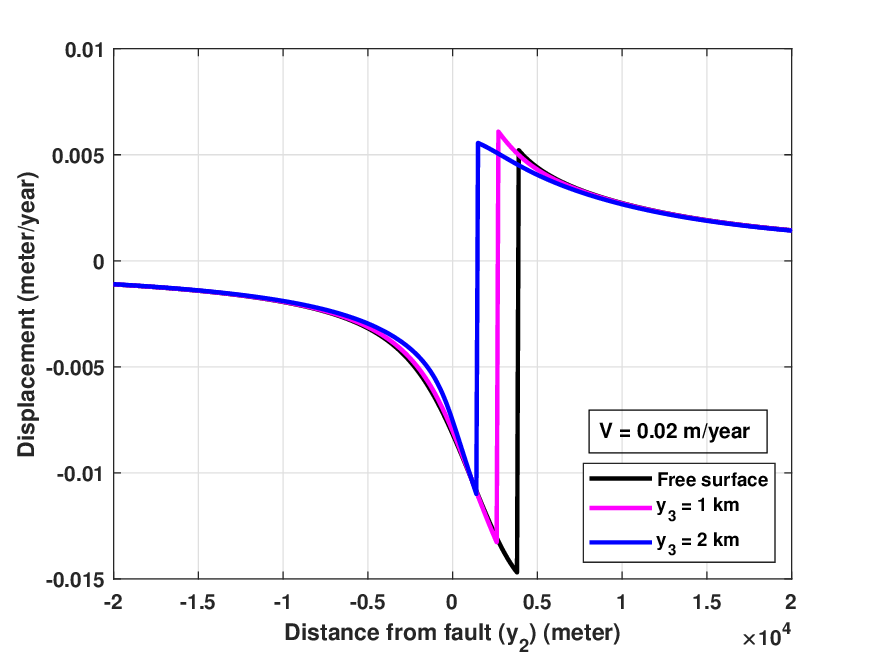}
  \caption{}
  \label{figure2b}
\end{subfigure}
\caption{Changes of displacement against $y_2$ (a) for different creep velocity $V$ (b) for different $y_3$.}
\label{figure2}
\end{figure}

\begin{figure}[ht!] 
\centering
\includegraphics[width=0.7\textwidth]{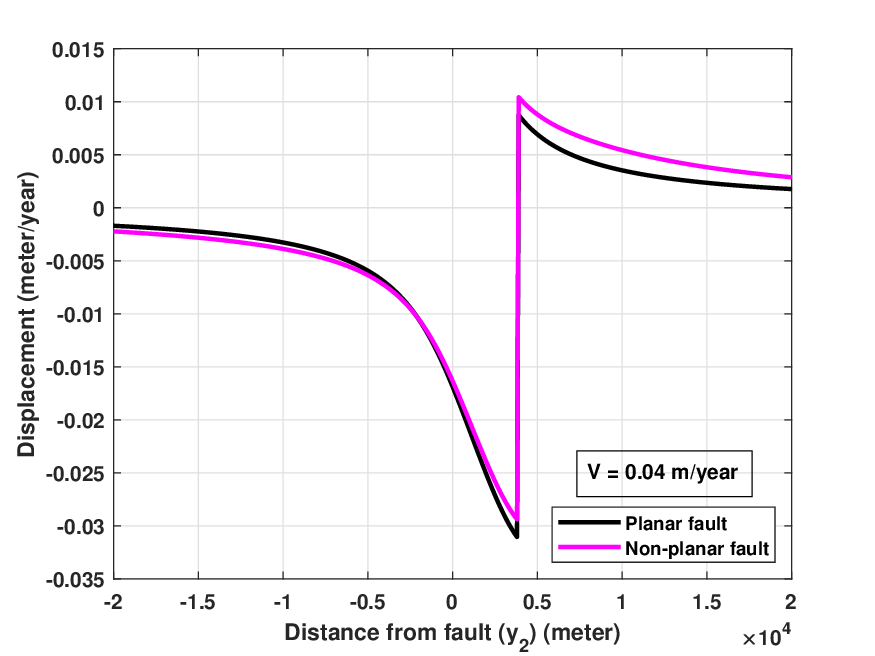}
    \caption{Change of surface displacement against $y_2$ for planar and non-planar fault.}
    \label{figure3}
\end{figure}

 Our numerical computations primarily aim to examine surface displacements, the evolution of stress accumulation and release, and the associated changes in strain distribution in the vicinity of the fault. The figures have been plotted using MATLAB R2022b.\\\\
\textbf{\textit{Analysis of surface displacement:}}\\\\
Initially, we determine the surface displacement resulting from fault motion after the system returns to its aseismic state. Figure \ref{figure2a} and \ref{figure2b} present the surface displacement $u_1$ against $y_2$ for different creep velocities $V$ and different $y_3$, respectively. The displacement component is found to be of the order of $10^{-3}$ against $y_2$, and its magnitude depends upon different creep velocities of the fault. The difference in the peak magnitude of displacement due to velocities is less pronounced for $y_2>0$ when compared to $y_2<0$. For $y_2>0$, the peak of the magnitude of displacement is maximum (approx. $0.01042$ m/year) for $V=0.04$ meter and is minimum (approx. $0.00521$ m/year) for $V=0.02$ meter. This finding is consistent with real-world observations; for instance, the displacement rate of the Kunlun and Awatere faults is approximately $0.0123-0.0066$ m/year as reported by Gold and Cowgill (2011). From figure \ref{figure2b}, it has been observed that the absolute magnitude of displacement is highest on the free surface, and gradually decreases with increasing depth of the field point. In both scenarios, the displacement magnitude progressively diminishes with increasing distance from the fault on both sides. Finally, as $|y_2| \to \infty$, the displacement approaches zero, indicating that far-field displacement is negligible. This behaviour is consistent with physical expectations and real-world observations (Segall 2010). It is to be noted that when such displacement is compared to that of the fault model proposed by Mahato et al. (2022), the displacement is found to be of the order of \(10^{-3}\) for both cases. However, the displacement patterns differ, likely due to the influence of the non-planar fault geometry and variations in the viscoelastic model used.
\\\\
Figure \ref{figure3} illustrates the variation in surface displacement on the free surface ($y_3=0$) against $y_2$ for planar and non-planar faults under a steady creep rate of $V=0.04$ m/year. For the planar fault, we have considered only the part $AB$ of our fault model. The tip point of both the faults is at a distance of approximately. $3830$ m from the origin ($y_2=y_3=0$) and the maximum displacement due to both planar and non-planar faults on the free surface is achieved approximately at a distance of $3900$ m, which is quite feasible from a physical point of view. Notably, the non-planar fault shows a sharper and slightly greater displacement, particularly in the vicinity of the fault trace, when compared to the planar fault model. The non-planarity causes asymmetry in the displacement profile, suggesting that the geometry of the fault plays a crucial role in controlling surface deformation patterns.

\begin{figure}[ht]
\begin{subfigure}{.5\textwidth}
  \centering
  \includegraphics[width=1\linewidth]{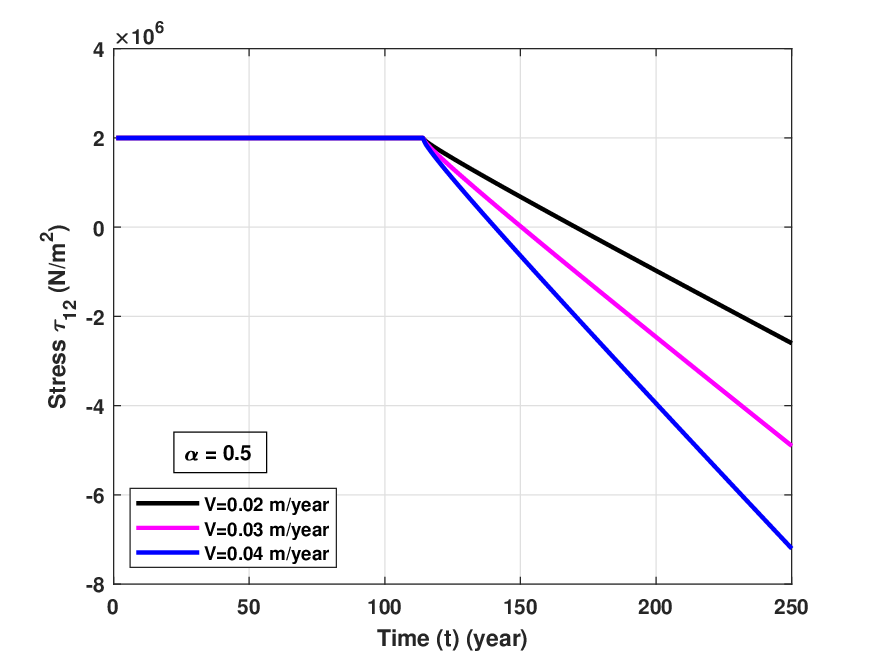}
  \caption{}
  \label{figure4a}
\end{subfigure}
\begin{subfigure}{.5\textwidth}
  \centering
  \includegraphics[width=1\linewidth]{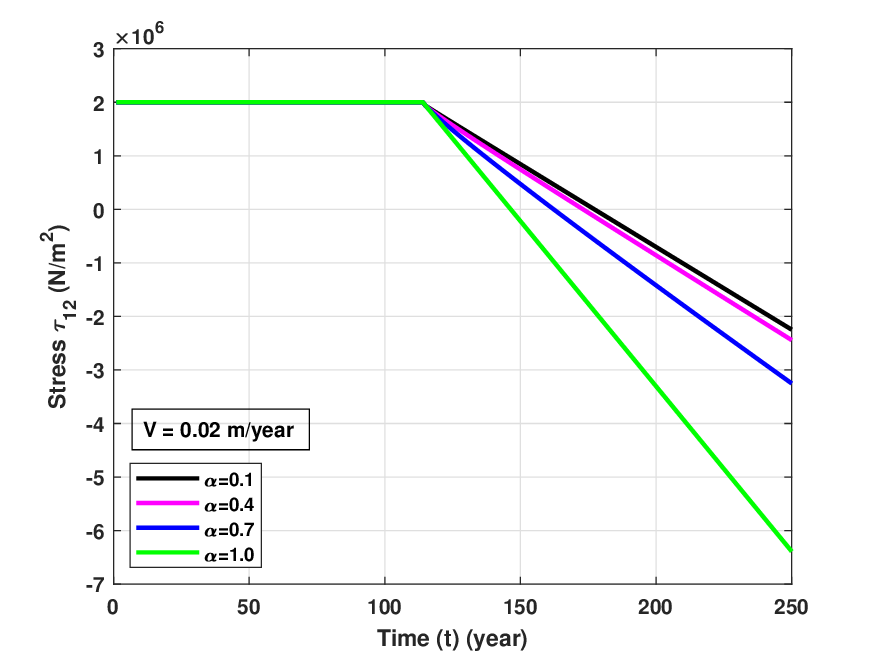}
  \caption{}
  \label{figure4b}
\end{subfigure}
\caption{Changes of surface stress $\tau_{12}$ against time $t$ for (a) different creep velocity $V$ (b) different order of the fractional derivative $\alpha$.}
\label{figure4}
\end{figure}

\begin{figure}[ht]
\begin{subfigure}{.5\textwidth}
  \centering
  \includegraphics[width=1\linewidth]{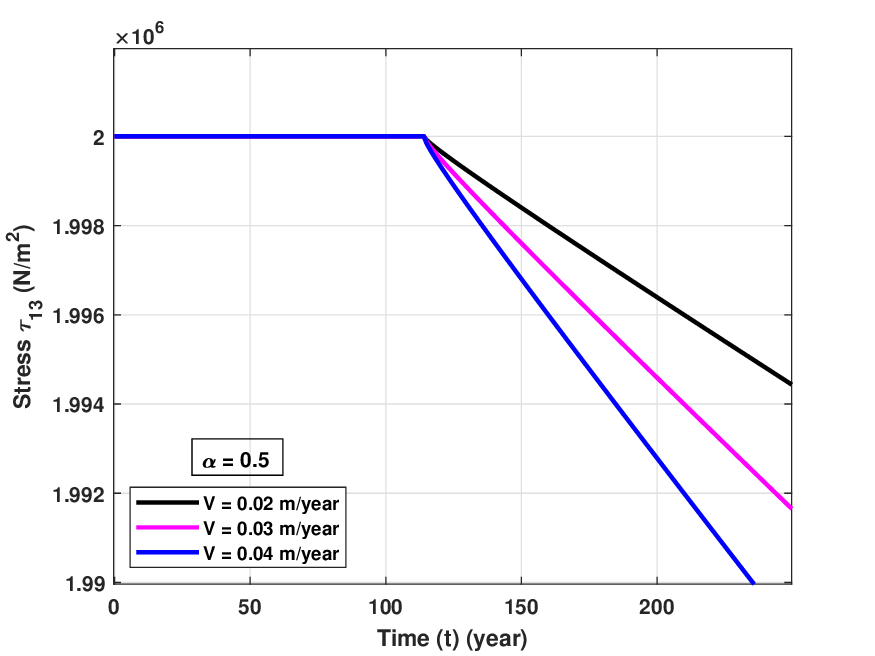}
  \caption{}
  \label{figure5a}
\end{subfigure}
\begin{subfigure}{.5\textwidth}
  \centering
  \includegraphics[width=1\linewidth]{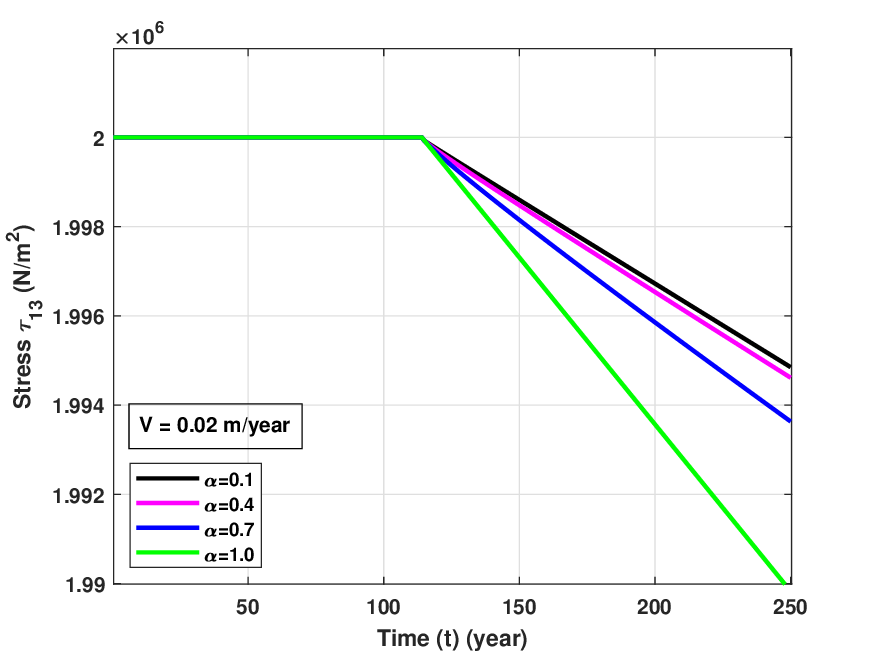}
  \caption{}
  \label{figure5b}
\end{subfigure}
\caption{Changes of stress $\tau_{13}$ against time $t$ for (a) different creep velocity $V$ (b) different order of the fractional derivative $\alpha$.}
\label{figure5}
\end{figure}

\begin{figure}[ht!] 
  \centering
\includegraphics[width=0.7\textwidth]{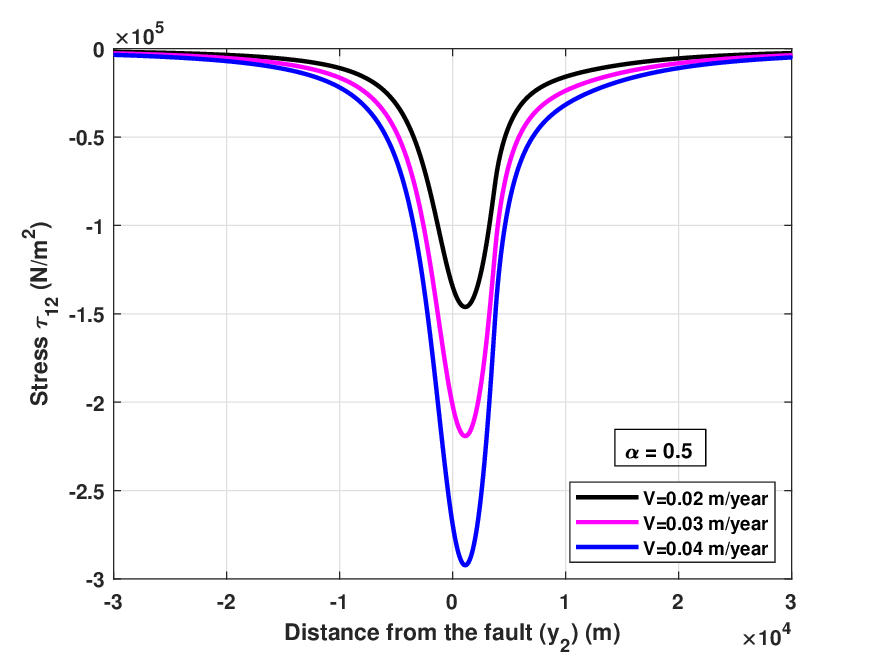}
    \caption{Changes of surface shear stress $\tau_{12}$ against $y_2$ for different creep velocity $V$.}
    \label{figure6}
\end{figure}

\begin{figure}[ht!]
  \centering
\includegraphics[width=0.7\textwidth]{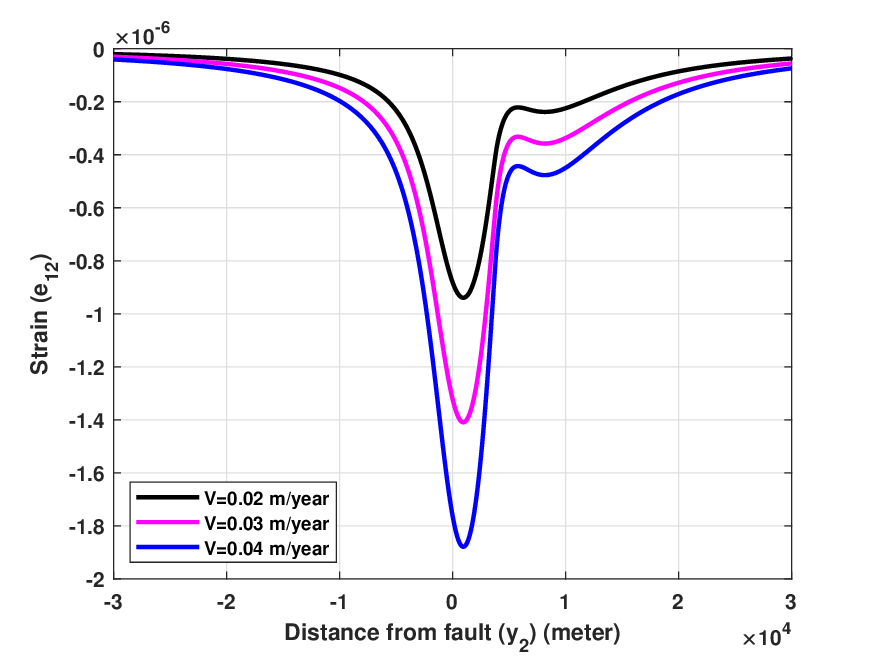}
    \caption{Change of surface shear strain $e_{12}$ against $y_2$ due to the movement across the fault.}
    \label{figure7}
\end{figure}

These results emphasize the importance of incorporating realistic fault geometries like non-planar faults in deformation models to better capture the spatial distribution of displacement, which is critical for geodetic interpretations and seismic hazard assessments.\\\\
\\
\textbf{{\textit{Analysis of stress accumulation/release}:}}\\\\
Figure \ref{figure4a} represents the changes of surface shear stress $\tau_{12}$ against time $t$ for different creep velocities $V$ taking a fixed $\alpha=0.5$. The stress starts to drop when observed at $t=115$ years after fault movement at $114.01$ years with different rates depending on the creep velocity. The shear stress $\tau_{12}$ when creep velocity $V$ is $0.04$ m/year (blue line) shows the fastest decrease in stress, while for $V=0.02$ m/year (black line), it shows comparatively slower decrease. Figure \ref{figure4b} illustrates the evolution of surface shear stress $\tau_{12}$ over time in a viscoelastic medium for different values of the fractional order $\alpha$, taking a fixed creep velocity $V=0.02$ m/year. A rapid stress relaxation has been shown for $\alpha=1.0$ while this relaxation is slowest for $\alpha=0.1$, i.e., the system exhibits a more substantial memory effect, leading to slower stress dissipation for decreasing value of $\alpha$. 
\\\\
Figure \ref{figure5a} and \ref{figure5b} depict the variation of surface shear stress $\tau_{13}$ over time $t$ for various creep velocities $V$ and distinct $\alpha$. It is observed that the stress starts to decline after around 114.01 years, although at a different rate compared to \(\tau_{12}\) (Figure \ref{figure4}). The rate at which \(\tau_{13}\) accumulates or dissipates is slower than that of \(\tau_{12}\). Our analysis reveals that variations in surface shear stress are influenced by the order of the fractional derivative $\alpha$. The resulting stress variations, which is of the order of $10^6$ $N/m^2$, are compatible with the range reported by Rafie et al. (2023), who observed values between $3\times10^5$ $N/m^2$ to $3.4\times 10^6$ $N/m^2$. This alignment strengthens the credibility of the fractional model and demonstrates its effectiveness in realistically capturing the observed stress variations.
\\\\
Figure \ref{figure6} illustrates the variation of surface shear stress $\tau_{12}$ against $y_2$ for different creep velocities $V$, taking fixed $\alpha=0.5$. As the creep velocity $V$ increases, the release of shear stress becomes larger. This implies that higher slip rates on the fault result in more extensive and intense stress changes in the surrounding medium. The stress release is highest near the fault. Moving away from the fault in either direction, the shear stress decreases rapidly and eventually stabilizes to nearly zero. This trend is physically reasonable, as shear stress tends to concentrate near the fault and relaxes away from it. The higher the slip velocity, the wider and taller the stress peak, suggesting that faster slip leads to a more significant stress perturbation in the surrounding rock. The stress distribution is symmetric about the fault, reflecting a balanced shear field on either side.\\
  \\
\textbf{{\textit{Analysis of strain}:}}\\\\
Figure \ref{figure7} illustrates the distribution of the strain $e_{12}$ against $y_2$ on the free surface for different $V$. Shear strain is highest near the fault, i.e., the deformation is most intense at the fault line. As the creep velocity increases, the peak strain magnitude also rises. Additionally, the surface shear strain exhibits a noticeable shift as the distance from the fault origin increases on either side, gradually approaching zero as $|y_2| \to \infty$. The magnitude of maximum strain for $V = 0.04$ m/year is approx. $1.879\times 10^{-6}$ while it is approx. $9.393\times 10^{-7}$ for $V=0.02$ m/year. The strain changes in the order of $10^{-6}$, which aligns with the finding of Takagi and Okubo (2017), which reports that the strain variations range from $10^{-4}$ to $10^{-7}$. As $y_2$ approaches infinity, the strain $e_{12}$ gradually diminishes to zero. In practical situations, strain typically decreases with increasing distance from the stress source, as noted by Segall (2010). This behaviour is consistent with both theoretical predictions and empirical observations. Compared to the model by Mahato et al. (2022), which considers an infinite planar fault in a standard linear solid viscoelastic medium, notable differences arise. The planar fault model estimates strain variations around the order of $10^{-7}$, whereas the non-planar fault model considered in this work shows strain changes reaching the order of $10^{-6}$. This discrepancy is primarily due to the impact of fault geometry and the application of distinct viscoelastic frameworks.    
\section{Conclusion \& future scope}\label{sec5}
This study examines a non-planar (with three planar cross-sectional parts) surface-breaking infinite strike-slip fault, which is embedded in a viscoelastic half-space of fractional Maxwell material. We analyze the impact of fault creep on displacement, stress, and strain. The order of the fractional derivative $\alpha$ is taken within the range $0 < \alpha \leq 1$. Figures \ref{figure2a} and \ref{figure2b} indicate that the displacement rates are approximately in the order of $10^{-3}$ m/year, closely matching the values observed in real-world studies, such as those by Lyons and Sandwell (2009).  

Additionally, figures \ref{figure4} and \ref{figure5} demonstrate that the build-up and release of shear stress vary with the creep velocity of the fault as well as the order of the fractional derivative. A comparison between figures \ref{figure4} and \ref{figure5} reveals that the shear stress component $\tau_{13}$ undergoes a greater release than $\tau_{12}$. The resulting displacement, stress, and strain distributions predicted by our model exhibit strong agreement with observed geological data, as discussed in the respective sections. Figures \ref{figure6} and \ref{figure7} demonstrate the significant impact of fault's creep velocity on surface shear stress and strain, respectively. As the creep velocity $V$ increases, both the stress and strain components near the fault intensify and extend over a broader region. Higher slip rates not only amplify the peak stress and strain but also result in more widespread deformation across the fault zone.
\\\\
Future research could focus on developing mathematical models that involve infinite interacting non-planar faults or combinations of finite and infinite non-planar fault systems. Furthermore, the medium may be represented as a viscoelastic half-space characterized by either fractional Burgers rheology or a fractional standard linear solid model. Additionally, modern machine learning and deep learning methods (Khatir et al., 2022, 2023) hold significant potential in structural monitoring and fault detection. These data-centric methods can serve as valuable tools alongside traditional models, improving the precision of stress distribution forecasting and the understanding of fault behavior. Future studies will focus on incorporating these techniques more extensively to enhance predictive modeling capabilities and deepen our insight into fault dynamics.
\section*{Declarations}
\begin{itemize}
\item \textbf{Acknowledgement}\\
The first author extends sincere gratitude to the National Institute of Technology, Durgapur, India, for providing both financial assistance and logistical support that facilitated the successful completion of this research.
\item \textbf{Conflict of interest}\\
The authors affirm that there are no conflicts that could have influenced the outcomes of this work.
\item \textbf{Availability of data and materials}\\
The study incorporates data sourced from previously published literature, with appropriate citations provided within the relevant sections of the manuscript.
\item \textbf{Code availability}\\
All computational works were conducted using MATLAB software, version R2022b.   
\item \textbf{Authors' contributions}\\
The first author was responsible for drafting the manuscript and was primarily involved in the development and execution of the computational model. The second author provided overall guidance, refined the manuscript, and contributed to improving its linguistic quality and structure.
\end{itemize}
\section*{References}
\begin{itemize}
\item[1.] Aochi H, Eiichi F, Matsu’ura M (2000) Spontaneous rupture propagation on a non-planar fault in 3-D elastic medium. Pure and Applied Geophysics 157:2003-2027
\item[2.] Bischoff SH, Flesch LM (2018) Normal faulting and viscous buckling in the Tibetan Plateau induced by a weak lower-crust. Nature communications 9(1):4952
\item[3.] Cruz-Atienza VM, Virieux J (2004) Dynamic rupture simulation of non-planar faults with a finite-difference approach. Geophysical Journal International 158(3):939-954 
\item[4.] Debnath L (2003) Recent applications of fractional calculus to science and engineering. International Journal of Mathematics and Mathematical Sciences 2003(54):3413–3442 
\item[5.] Deng W, Morozov IB (2018) Mechanical interpretation and generalization of the Cole–Cole model in viscoelasticity. Geophysics 83(6):345–352
\item[6.] El-Misiery AEM, Ahmed A (2006) On a fractional model for earthquakes. Applied mathematics and computation 178(2):207-211. https://doi.org/10.1016/j.amc.2005.10.011
\item[7.] Folesky J (2024) Different earthquake nucleation conditions revealed by stress drop and b-value mapping in the northern Chilean subduction zone. Scientific Reports 14(1):12182 
\item[8.] Ge W, Shen Z, Molnar P, Wang M, Zhang P, Yuan D (2022) GPS determined asymmetric deformation across central Altyn Tagh fault reveals rheological structure of northern Tibet. Journal of Geophysical Research: Solid
Earth 127:e2022JB024216. https://doi.org/10.1029/2022JB024216
\item[9.] Gold RD, Cowgill E (2011) Deriving fault-slip histories to test for secular variation in slip, with examples from the Kunlun and Awatere faults. Earth and Planetary Science Letters 301(1-2):52-64. https://doi.org/10.1016/j.epsl.2010.10.011
\item[10.] Hisakawa T, Ando R, Yano TE, Matsubara, M (2020) Dynamic rupture simulation of 2018, Hokkaido Eastern Iburi earthquake: role of non-planar geometry. Earth, Planets and Space. 72:1-14. https://doi.org/10.1186/s40623-020-01160-y
\item[11.] Khatir A, Capozucca R, Khatir S, Magagnini E (2022) Vibration-based crack prediction on a beam model using hybrid butterfly optimization algorithm with artificial neural network. Frontiers of Structural and Civil Engineering 16(8):976–989. https://doi.org/10.1007/s11709-022-0840-2
 \item[12.] Khatir A, Capozucca R, Khatir S, Magagnini E, Benaissa B, Thanh CL, Wahab MA (2023) A new hybrid PSO-YUKI for double cracks identification in CFRP cantilever beam. Composite Structures 311:116803. https://doi.org/10.1016/j.compstruct.2023.116803 
\item[13.] Koller MG, Bonnet M, Madariaga R (1992) Modeling of dynamical crack propagation using time-domain boundary integral equations. Wave Motion 16(4):339–366. https://doi.org/10.1016/0165-2125(92)90022-T
\item[14.] Li Q, Liu M, Zhang H (2009) A 3-D viscoelastoplastic model for simulating long-term slip on non-planar faults. Geophysical Journal International 176(1):293-306.  https://doi.org/10.1111/j.1365-246X.2008.03962.x
 \item[15.] Lyons S, Sandwell D (2003) Fault creep along the southern San Andreas from interferometric synthetic aperture radar, permanent scatterers, and stacking. Journal of Geophysical Research: Solid Earth 108(B1).  https://doi.org/10.1029/2002JB001831
 \item[16.] Mahato P, Mondal D, Sarkar (Mondal) S (2022) Determination of effect of the movement of an infinite fault in viscoelastic half space of standard linear solid using fractional calculus. Physica Scripta 97(12):125015.
 https://doi.org/10.1088/1402-4896/ac9caa
 \item[17.] Mahato P, Sarkar Mondal S (2025) Effect on surface deformation due to interacting fault in fractional standard linear solid. GEM-International Journal on Geomathematics 16(1):5. https://doi.org/10.1007/s13137-025-00264-5 
 \item[18.] Mondal D, Debnath P (2021) An application of fractional calculus to geophysics: effect of a strike-slip fault on displacement, stresses and strains in a fractional order Maxwell type visco-elastic half space. International Journal of Applied Mathematics 34(5):873. https://doi.org/10.12732/ijam.v34i5.2
 \item[19.] Mondal SC, Debsarma S (2023) Numerical modelling of a nonplanar strike slip fault and associated stress distribution in lithosphere asthenosphere system. GEM-International Journal on Geomathematics 14(1):15.  https://doi.org/10.1007/s13137-023-00222-z
\item[20.] Mukhopadhyay A, Sen S, Pal BP (1980) On stress accumulating in a viscoelastic lithosphere containing a continuously slipping fault. Bulletin Society of Earthquake Technology 17(1):1–10
 \item[21.] Oglesby DD, Mai PM (2012) Fault geometry, rupture dynamics and ground motion from potential earthquakes on the North Anatolian Fault under the Sea of Marmara. Geophysical Journal International 188(3):1071–1087. https://doi.org/10.1111/j.1365-246X.2011.05289.x 
 \item[22.] Pelap FB, Tanekou GB, Fogang CF, Kengne R (2018) Fractional-order stability analysis of earthquake dynamics. Journal of Geophysics and Engineering 15(4):1673–1687. https://doi.org/10.1088/1742-2140/aabe61
 \item[23.] Pelties C, Puente J, Ampuero JP, Brietzke GB, Kaser M (2012) Three-dimensional dynamic rupture simulation with a high-order discontinuous Galerkin method on unstructured tetrahedral meshes. Journal of Geophysical Research: Solid Earth 117:B02309. https://doi.org/10.1029/2011jb008857
 \item[25.] Rafie MT, Sahara DP, Cummins PR, Triyoso W, Widiyantoro S (2023) Stress accumulation and earthquake activity on the Great Sumatran Fault, Indonesia. Natural Hazards. 116(3):3401–3425. https://doi.org/10.1007/s11069-023-05816-2 
 \item[26.] Romanet P, Saito T, Fukuyama E (2024) The mechanics of static non-planar faults in infinitesimal strain theory. Geophysical Journal International 239(3):1664-1693. https://doi.org/10.1093/gji/ggae337
 \item[27.] Samko SG, Kilbas AA, Marichev OI (1993) Fractional Integrals and Derivatives: Theory and Applications. CRC, New York (1993)
 \item[28.] Seelig Th, Gross D (1997) Analysis of dynamic crack propagation using a time-domain boundary integral equation method. International Journal of Solids and Structures 34(17):2087–2103 (1997). https://doi.org/10.1016/S0020-7683(96)00133-3
 \item[29.] Segall P (2010) Earthquake and Volcano Deformation. Earthquake and Volcano Deformation, Princeton University Press 
 \item[30.] Tada T, Yamashita T (1996) The paradox of smooth and abrupt bends in two-dimensional in-plane shear-crack mechanics. Geophysical Journal International 127(3):795–800.  https://doi.org/10.1111/j.1365-246X.1996.tb04058.x
  \item[31.] Takagi Y, Okubo S (2017) Internal deformation caused by a point dislocation in a uniform elastic sphere. Geophysical Journal International 208(2):973–991 (2017). https://doi.org/10.1093/gji/ggw424 
 \item[32.] Tanekou GB, Fogang CF, Pelap FB, Kengne R, Fozin TF, Nbendjo BRN (2020) Complex dynamics in the two spring-block model for earthquakes with fractional viscous damping. The European Physical Journal Plus 135(7):1-26. https://doi.org/10.1140/epjp/s13360-020-00558-7
 \item[33.] Titus SJ, DeMets C, Tikoff B (2006) Thirty-five-year creep rates for the creeping segment of the San Andreas fault and the effects of the 2004 Parkfield earthquake: constraints from alignment arrays, continuous global positioning system, and creepmeters. Bulletin of the Seismological Society of America 96(4B):S250–S268. https://doi.org/10.1785/0120050811 
 \item[34.] Turcotte DL, Gerald S (2002) Geodynamics. Cambridge University Press, Cambridge
 \item[35.] Wu F, Ji C, Liu J, Gao R, Li C, Zou Q, Chen J (2023) Study on visco-elastoplastic fractional creep model of surrounding rock of salt cavern gas storage. Journal of Energy Storage 67:107606. https://doi.org/10.1016/j.est.2023.107606
  \item[36.] Zielke O, Mai PM (2025) Does Subsurface Fault Geometry Affect Aleatory Variability in Modeled Strike-Slip Fault Behavior?. Bulletin of the Seismological Society of America 115(2):399–415. https://doi.org/10.1785/0120240152  
\end{itemize}

 


\end{document}